\documentclass{ipa2e}      
\usepackage{epsfig}        
\usepackage{html}
\Title{Multi-step VLBI observations of weak extragalactic radio sources to align the ICRF and the future GAIA frame}
\STitle{VLBI survey \& ICRF--GAIA link}
\PublicationType{proceedings}
\PublicationName{IVS 2008 General Meeting Proceedings}
\IvsComponentName{Bordeaux Observatory Analysis Center}
\NumberofAuthors{4}
\AuthorName{1}{G\'{e}raldine Bourda}
\AuthorName{2}{Patrick Charlot}
\AuthorName{3}{Richard Porcas}
\AuthorName{4}{Simon Garrington}
\ContactAuthorReference{1}
\AuthorEmail{1}{bourda@obs.u-bordeaux1.fr}
\AuthorTelephone{1}{+33-5-57-77-61-27}
\AuthorPersonalWebPage{1}{}     
\AuthorEmail{2}{}
\AuthorTelephone{2}{}
\AuthorPersonalWebPage{2}{}     
\AuthorEmail{3}{}
\AuthorTelephone{3}{}
\AuthorPersonalWebPage{3}{}     
\AuthorEmail{4}{}
\AuthorTelephone{4}{}
\AuthorPersonalWebPage{4}{}  
\AuthorInstitutionReference{1}{1}
\AuthorInstitutionReference{2}{1}
\AuthorInstitutionReference{3}{2}
\AuthorInstitutionReference{4}{3}
\NumberofInstitutions{3}
\InstitutionName{1}{Laboratoire d'Astrophysique de Bordeaux, Universit\'{e} Bordeaux~1, CNRS}
\InstitutionPostAddress{1}{BP89, 33270 Floirac}
\InstitutionCountry{1}{France}
\InstitutionFax{1}{+33-5-57-77-61-10}
\InstitutionWebPage{1}{http://www.obs.u-bordeaux1.fr}
\InstitutionName{2}{Max Planck Institute for Radio Astronomy, Bonn}
\InstitutionPostAddress{2}{P.O. Box 20 24, 53010 Bonn}
\InstitutionCountry{2}{Germany}
\InstitutionFax{2}{}
\InstitutionWebPage{2}{http://www.mpifr-bonn.mpg.de}
\InstitutionName{3}{Jodrell Bank Observatory, The University of Manchester}
\InstitutionPostAddress{3}{Macclesfield, Cheshire, SK11 9DL}
\InstitutionCountry{3}{UK}
\InstitutionFax{3}{}
\InstitutionWebPage{3}{http://www.jb.man.ac.uk}
\begin{document}
\noindent\mbox{\small The $5^{th}$ IVS General Meeting Proceedings, 2008, p.?--?}
\vskip6mm\centerline{\large Section name}\vskip3mm\hrule height1pt\vskip10mm

\MakeTitle                

\begin{abstract}
The space astrometry mission GAIA will construct a dense optical QSO-based celestial reference frame. For consistency between optical and radio positions, it will be important to align the GAIA frame and the International Celestial Reference Frame (ICRF) with the highest accuracy. Currently, it is found that only 10\% of the ICRF sources are suitable to establish this link, either because they are not bright enough at optical wavelengths or because they have significant extended radio emission which precludes reaching the highest astrometric accuracy. In order to improve the situation, we have initiated a VLBI survey dedicated to finding additional suitable radio sources for aligning the two frames. The sample consists of about 450 sources, typically 20 times weaker than the current ICRF sources (down to the 20 mJy flux level), which have been selected by cross-correlating optical and radio catalogues. This paper presents the observing strategy to detect, image, and measure accurate positions for these sources. It will also provide results about the VLBI detectability of the sources, as derived from initial observations with the European VLBI Network in June and October 2007. Based on these observations, an excellent detection rate of 89\% is found, which is very promising for the continuation of this project.
\end{abstract}

\section{Context}

The ICRF (International Celestial Reference Frame; \cite{Ma1998} \cite{Fey2004}) is the fundamental celestial reference frame adopted by the International Astronomical Union (IAU) in August 1997. It is currently based on the VLBI (Very Long Baseline Interferometry) positions of 717 extragalactic radio sources, estimated from dual-frequency S/X (2.3 and 8.6 GHz) observations. 
The European space astrometry mission GAIA, to be launched by 2011, will survey about one billion stars in our Galaxy and 500~000 Quasi Stellar Objects (QSOs) brighter than magnitude 20 \cite{Perryman2001}. Unlike Hipparcos, GAIA will construct a dense optical celestial reference frame directly at optical bands, based on the QSOs with the most accurate positions (i.e. those with optical apparent magnitude $V\leq18$ \cite{Mignard2003}). 
In the future, the alignment of the ICRF and the GAIA frame will be crucial, in particular for ensuring consistency between measured radio and optical positions. This alignment, to be determined with the highest accuracy, requires hundreds of sources in common, with a uniform sky coverage and very accurate radio and optical positions. Obtaining such accurate positions implies that the link sources must have $V\leq18$ and no extended VLBI structures. 

In a previous study, we investigated the current status of this link based on the present list of ICRF sources \cite{Bourda2008}. We found that although about 30\% of the ICRF sources have an optical counterpart with $V \leq 18$, only one third of these are compact enough on VLBI scales for the highest astrometric accuracy. 
Overall, only 10\% of the current ICRF sources (70~sources) are thus available for the alignment with the GAIA frame. This highlights the need to identify additional suitable radio sources, which is the purpose of the project described here.

\section{Strategy to identify new link radio sources}

Searching for additional radio sources suitable for aligning accurately the ICRF and the GAIA frame could rely on the VLBA Calibrator Survey (VCS; \cite{Petrov2008} and references therein), the catalogue of more than 3000 extragalactic radio sources observed with the VLBA (Very Long Baseline Array; the american VLBI network), which is currently underway. Another possibility is to search for new VLBI sources, which implies going to weaker radio sources with flux densities typically below 100~mJy. This can now be realized owing to recent increases in the VLBI network sensitivity (e.g. recording now possible at 1Gb/s) and by using a network comprising large antennas like the European VLBI Network (EVN).

A sample of about 450 radio sources, for which there are no published VLBI observations, was selected for this purpose by cross-identifying the NRAO VLA Sky Survey (NVSS \cite{Condon1998}), a deep radio survey (complete to the 2.5~mJy level) which covers the entire sky north of $-40^{\circ}$, with the V\'{e}ron~\&~V\'{e}ron~(2006) optical catalogue \cite{Veron2006}. This sample is based on the following criteria: $V\leq18$ (to ensure very accurate positions with GAIA), $\delta\geq-10^{\circ}$ (for possible observing with northern VLBI arrays), and NVSS flux density~$\geq$~20~mJy (for possible VLBI detection). 

The observing strategy to identify the most appropriate link sources from this sample includes three successive steps to detect, image and measure accurate positions for these sources: (i) to determine the VLBI detectability of these weak radio sources; (ii) to image the sources detected in the first step, in order to identify the most point-like sources; and (iii) to determine an accurate astrometric position for the most compact sources of the sample.

\section{Initial VLBI results}

Initial VLBI observations for this project were carried out in June and October 2007 (during two 48-hours experiments, named EC025A and EC025B, respectively), with a network of four EVN telescopes (Effelsberg, Medicina, Noto, Onsala; and the 70~m Robledo telescope for part of the time in EC025B). The purpose of these two experiments was to determine the VLBI detectability of the 447 weak radio sources in our sample based on snapshot observations. 

\begin{figure}[htb!]
\begin{center}
\epsfig{file=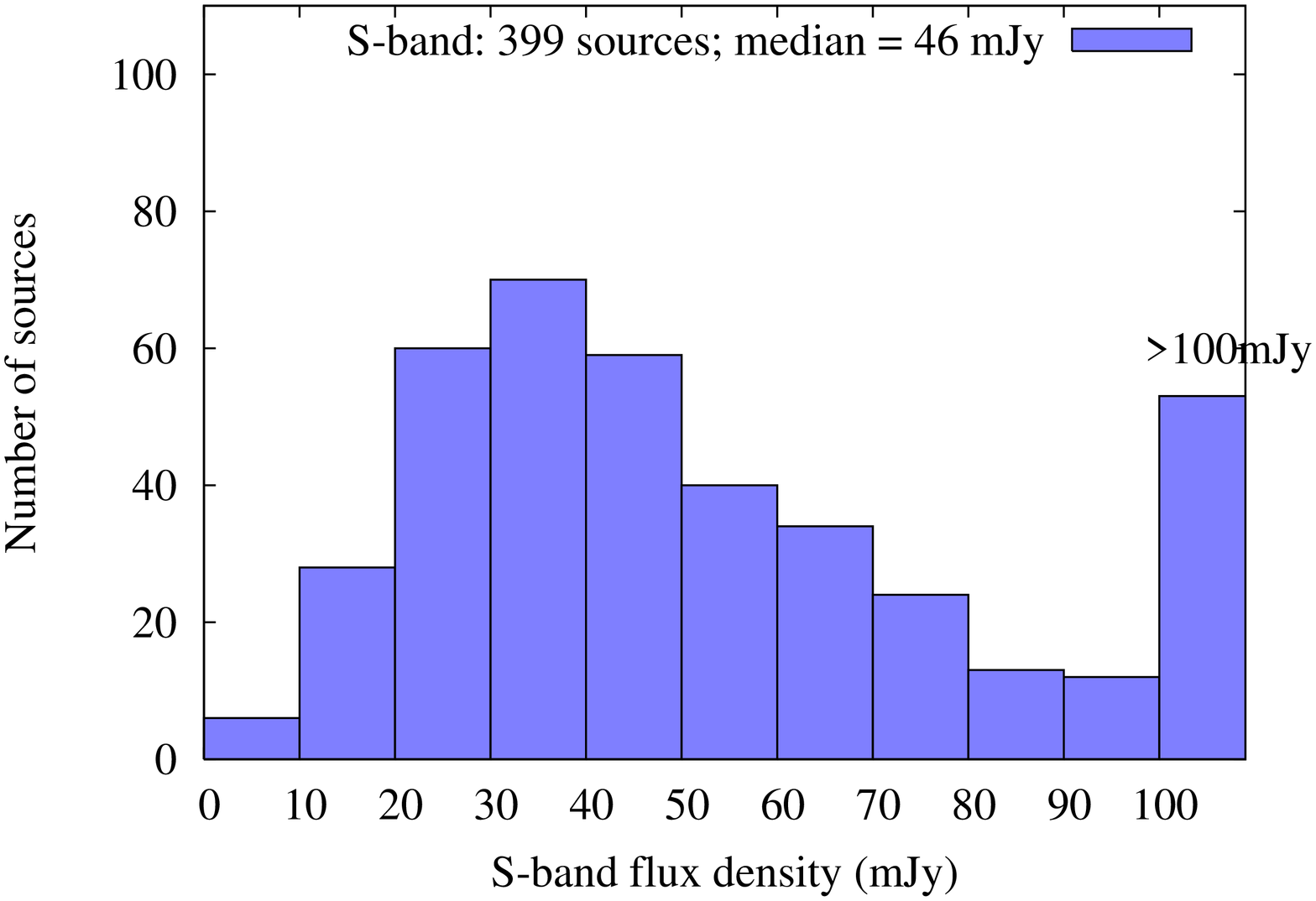,scale=0.26,angle=-90} 
\epsfig{file=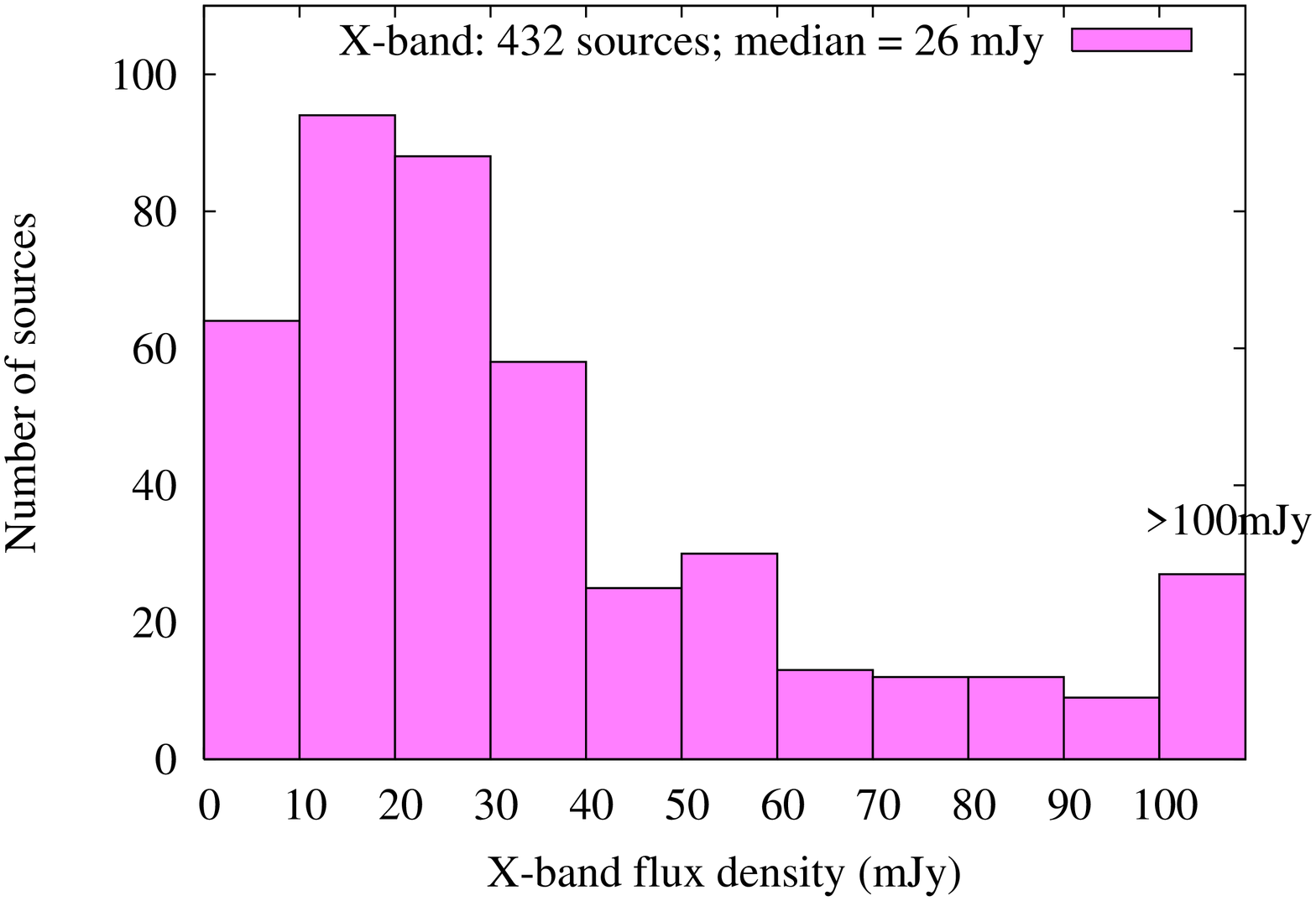,scale=0.26,angle=-90} 
\end{center}
\caption{Mean correlated flux density distribution (units in mJy), at S band and X band, for the sources detected in EC025A and EC025B, conducted respectively in June and October 2007.}
\label{fig:Fig1}
\end{figure}

Our results indicate excellent detection rates of 97\% at X band and 89\% at S band, with 432~sources and 399~sources detected at X and S bands, respectively (the individual results for EC025A and EC025B are detailed separately in Table~\ref{tab:Tab1}). This detection rate is in agreement with that reported in \cite{Frey2008} (80\%) for quasars from the Sloan Digital Sky Survey. The overall mean correlated flux densities (i.e. for each source and band, the mean over all scans and baselines detected) have a median value of 26~mJy at X~band and 46~mJy at S~band (see Figure~\ref{fig:Fig1}) with the weakest sources at the level of 1~mJy at X~band and 8~mJy at S~band. A comparison between the X-band flux density distribution for our sources, those from the VCS and the ICRF sources shows that the sources of our sample are indeed much weaker. On average, they are 27~times weaker than the ICRF sources and 8~times weaker than the VCS sources.  

\begin{figure}[htb!]
\begin{center}
\epsfig{file=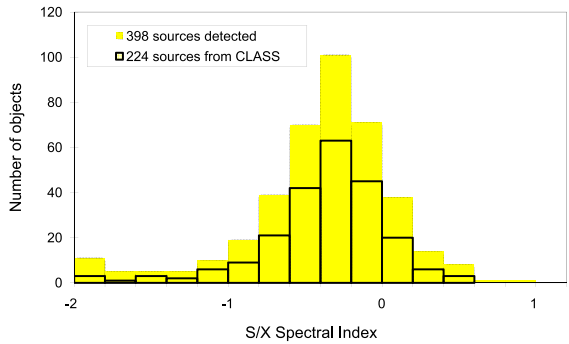,scale=0.31,angle=-90} 
\end{center}
\caption{S/X spectral index distribution for the 398 weak extragalactic radio sources detected at both S and X bands during the experiments EC025A and EC025B. Additionally, the S/X spectral index distribution for the sources also belonging to the CLASS catalogue is plotted (in black).}
\label{fig:Fig2}
\end{figure}

The spectral index $\alpha$ (defined as $S \propto \nu^{\alpha}$, where $S$ is the source flux density and $\nu$ is the frequency) was also investigated. In this definition, the sources with a compact core are expected to have $\alpha > -0.5$. Figure~\ref{fig:Fig2} shows the spectral index distribution for the 398 radio sources detected at both frequencies in EC025A and EC025B. The corresponding distribution for the sources which also belong to the CLASS catalogue \cite{Myers2003}, well known to be composed of compact sources, is also plotted and no major differences are noticed. The median value of $\alpha$ is $-0.34$ and about 70\% of the sources have $\alpha > -0.5$, hence indicating that they must have a dominating core component, which is very promising for the future stages of this project.
\begin{table}[htb!] 
\caption{VLBI detection rate for the 447 weak extragalactic radio sources observed during EC025A and EC025B.} 
\label{tab:Tab1} 
\begin{center}     
\begin{tabular}{|l|c|c|c|c|} 
\hline
\hline
                & \textbf{Sources}  & \textbf{X-band}    & \textbf{S-band}    & \textbf{S and X}  \\
                & \textbf{observed} & \textbf{detection} & \textbf{detection} & \textbf{detection}\\
\hline
\textbf{EC025A} & 218               &   216 sources      &   211 sources      &  211 sources      \\  
                &                   &         99\%       &         96\%       &       96\%        \\ 
\hline
\textbf{EC025B} & 229               &   216 sources      &   188 sources      &  187 sources      \\ 
                &                   &         94\%       &         82\%       &       82\%        \\ 
\hline   
\textbf{Overall}& 447               &   432 sources      &   399 sources      &  398 sources      \\
                &                   &         97\%       &         89\%       &       89\%        \\ 
\hline   
\end{tabular}
\end{center}
\end{table}

\section{Summary}

Based on observations with the European VLBI Network, we identified 398 new VLBI sources which are potential candidates to align the ICRF and the future GAIA frame. On average, these sources are 27 times weaker than the ICRF sources. Overall, this multiplies by a factor of 6 the current number of potential ICRF--GAIA link sources (awaiting for the potential candidates identified from the VCS catalogue). 
The excellent detection rate inferred from the observations may suggest that our initial VLBI detection step is unnecessary for such radio sources having an optical counterpart with magnitude brighter than 18. 
Future steps will be targeted at imaging the 398 sources that we have detected at both frequencies by using the global VLBI network (EVN+VLBA), in order to identify the most point-like sources and therefore the most suitable ones for the ICRF--GAIA link.

\section{Acknowledgements}

The authors wish to thank Dave Graham for assistance with the correlation in Bonn, John Gipson for advice when scheduling the observations, and Alexander Andrei for providing improved optical positions. This work has benefited from research funding from the European Community's sixth Framework Programme under RadioNet R113CT 2003 5058187. The EVN is a joint facility of European, Chinese, South African and other radio astronomy institutes funded by their national research councils.


\end{document}